\documentclass[12pt, a4paper]{article} 

\usepackage[includeheadfoot,
            marginratio={1:1,2:3}, 
            width=452pt, 
            height=720pt,]{geometry}

\usepackage{amsmath}
\usepackage{amsfonts}
\usepackage{amssymb}
\usepackage{amsthm}
\usepackage{bbm,bm}
\usepackage{mathtools, comment}
\usepackage{slashed}
\usepackage{mathalfa}
\usepackage{graphicx,caption,subfigure}
\usepackage{tikz} 
\usepackage{tikz-cd}
\usetikzlibrary{decorations.pathreplacing,calc,tikzmark}
\usepackage{booktabs}
\usepackage{adjustbox}
\usepackage[utf8]{inputenc}
 \usepackage[splitrule]{footmisc}

\usepackage{placeins}
\usepackage{empheq}
\usepackage{paralist}
\usepackage{cite}
\usepackage[normalem]{ulem}
\usepackage{soul}

\usepackage[colorlinks=false,urlbordercolor=red
]{hyperref}
\usepackage{xcolor}
\usepackage{tensor}

\newcommand{\nc}{\newcommand}
\nc{\lb}{\llbracket}
\nc{\rb}{\rrbracket}
\nc{\gl}{\llbracket}
\nc{\gr}{\rrbracket}
\nc{\del}{\partial}
\nc{\tri}{\hspace{-3.5pt}\vartriangle\hspace{-3.5pt}}
\nc{\blacktri}{\blacktriangle}

\definecolor{rossoCP3}{cmyk}{0,.88,.77,.40}

\nc{\eq}[1]{\begin{equation}
                     \begin{split} #1 \end{split}
                     \end{equation}}
\nc{\ov}{\overline}

\nc{\fa}{\hat}
\nc{\fb}{\MakeUppercase}
\nc{\fc}{\tilde }
\nc{\Lie}{{\cal L}} 
\nc{\lambdabar}{{\mkern0.75mu\mathchar '26\mkern -9.75mu\lambda}}

\allowdisplaybreaks[2]
\numberwithin{equation}{section}

\DeclareUnicodeCharacter{2212}{-}
\begin{document}

\vspace*{-1.5cm}
\begin{flushright}
  {\small
  LMU-ASC 18/23\\ MPP-2023-97
  }
\end{flushright}

\vspace{1.5cm}
\begin{center}
  {\Large \bf Species Entropy and Thermodynamics} 
\vspace{0.35cm}

\end{center}

\vspace{0.35cm}
\begin{center}
{\large 
Niccol\`o Cribiori$^a$, Dieter L\"ust$^{a,b}$ and Carmine Montella$^{a,c}$
}
\end{center}

\vspace{0.1cm}
\begin{center} 
\emph{
$^a$Max-Planck-Institut f\"ur Physik (Werner-Heisenberg-Institut), \\[.1cm] 
   F\"ohringer Ring 6,  80805 M\"unchen, Germany, 
   \\[0.1cm] 
 \vspace{0.3cm}
             $^b${\it Arnold Sommerfeld Center for Theoretical Physics,\\
Ludwig-Maximilians-Universit\"at M\"unchen, 80333 M\"unchen, Germany},   \\[0.1cm] 
 \vspace{0.3cm}
$^c$Dipartimento di Fisica e Astronomia, Universit\`a di Bologna, \\
via Irnerio 46, 40126 Bologna, Italy. \\[.1cm] 
    } 
\end{center} 

\vspace{0.5cm}

\begin{abstract}

We analyse particle species and the species scale in quantum gravity from a thermodynamic perspective. 
In close analogy to black hole thermodynamics, we propose
that particle species have an entropy and a temperature, which is determined by the species scale. This is identical to the Bekenstein-Hawking entropy of a corresponding minimal black hole and agrees with the number of species in a given tower of states.  
Through the species entropy, we find that certain entropy bounds are connected to recent swampland constraints. 
Moreover, the concept of species entropy and temperature allow us to formulate the laws of species thermodynamics, which are argued to govern the variations of moduli in string theory.
They can be viewed as general rules that 
imply certain swampland conjectures, and vice versa.

\end{abstract}

\thispagestyle{empty}
\clearpage

\setcounter{tocdepth}{2}

\tableofcontents


\section{Introduction}

As is known from the seminal work of Bekenstein \cite{Bekenstein:1973ur} and Hawking \cite{Hawking:1975vcx,Hawking:1976de}, black holes are thermodynamic objects. 
They carry entropy, ${\cal S}_{BH}$, proportional to the area of their event horizon of size $R_{BH}$,
\begin{equation}
{\cal S}_{BH}\simeq(R_{BH})^{d-2}M_P^{d-2}\, ,\label{BHentropy}
\end{equation}
where $M_P$ is the Planck mass in $d$ space-time dimensions. 
They possess temperature, $T_{BH}$, determined by their surface gravity and, in semiclassical approximation, namely in the limit of large entropy, they emit a thermal radiation characterized by $T_{BH}$.
Black hole thermodynamics can be phrased in such a way that the well-known laws governing thermodynamic systems determine the variations of the black hole event horizon in General Relativity. 
This indicates that $\mathcal{S}_{BH}$ possesses a statistical interpretation in terms of so-called black hole microstates.

On the other hand, black holes are solutions of the Einstein equations in General Relativity and describe specific space-time metrics with an event horizon. 
This suggests that space-time itself possesses thermodynamic properties such as  entropy, and that gravity is an entropic force, as discussed \cite{Jacobson:1995ab,Verlinde:2010hp}. 
The notion of entropy of space-time also occurs e.g.~in the context of holography, namely in terms of the entanglement entropy between two regions in space-time \cite{Ryu:2006ef},
and also in the context of the ER=EPR proposal \cite{Maldacena:2013xja}.
Moreover, there are known bounds restricting the maximal amount of entropy allowed within a certain region of space-time.

The black hole horizon gives an intuitive notion of cutoff: what lies in the interior cannot be accessed directly from the outside, but it may still have interesting and non-trivial implications. In this work, we make this intuition more precise for what concerns the ultraviolet (UV) cutoff of gravitational effective theories. We accomplish this by proposing an analogy between black hole thermodynamics and certain properties of the spectrum of gravitational effective theories coupled to a high number of states.

Concretely, we consider a number of particles species coupled to gravity and we assume that their dynamics can be described by a $d$-dimensional effective action valid up to an UV cutoff scale $\Lambda_{UV}$. An upper bound for $\Lambda_{UV}$ is given by the so-called species scale $\Lambda_{sp}$, defined as the scale at which gravitational interactions become strong. 
For a $d$-dimensional theory with a number $N_{sp}$ of species with masses below $\Lambda_{sp}$, the species scale is given as \cite{Dvali:2007hz,Dvali:2007wp,Dvali:2009ks,Dvali:2010vm,Dvali:2012uq}
\begin{equation}\label{oldSC}
\Lambda_{sp} = \frac{M_{P}}{N_{sp}^{\frac{1}{d-2}}}
\end{equation}
and, compared to $M_P$, it is becoming lower and lower as more and more species are coupled to gravity.

Notice that, in general, $\Lambda_{sp}$ can be taken as a proxy for the UV cutoff of effective theories coupled to gravity. Of course, there can be other scales at which the effective description changes its nature and new particles have to be included. For example, when considering Kaluza-Klein (KK) compactifications from $D$ down to $d$ space-time dimensions, above a certain scale $m_{KK}$ determined by the radius $R_{KK}\simeq m_{KK}^{-1}$ of the compact space, a tower of KK particles enters the effective field theory, meaning that the latter becomes $D$-dimensional. However, even in this case the species scale remains the UV cutoff, in the sense that above $\Lambda_{sp}$ the effective description breaks down completely.

A high number of particle species occurs naturally in quantum gravity. In particular, in string theory there can be typically two towers of states becoming light at the boundary of the moduli space: string or KK states. According to the emergent string conjecture \cite{Lee:2019wij}, these are the only two possibilities in quantum gravity. Towers of string states, or better excitations of tensionless strings, become light at weak string coupling, or in case of S-duality at strong string coupling. Their number of species is determined by the string coupling constant $g_s$ \cite{Dvali:2009ks,Dvali:2010vm},
\begin{equation}
N_{sp}\simeq {\frac{1}{ g_s^2}}\, .\label{gs}
\end{equation}
Towers of KK states become light in the limit of decompactification to higher dimensions. Their number of species is given by
\begin{equation}
N_{sp}\simeq {\cal V} \simeq N_{KK}\, ,\label{vol}
\end{equation}
where ${\cal V}$ is the volume of the compact space.
When T-duality is considered, also towers of winding strings can be relevant and they become light in the limit of small internal volume.
Besides KK towers or string excitations, also $N_0$ massless or light particles with masses $m_0 \ll \Lambda_{sp}$ contribute to the species scale and lower the UV cutoff compared to $M_P$.
Note that the case of species that are massless or light in the entire moduli space is conceptually different and not continuously connected to that of tower of states becoming massless only at the boundary of the moduli space.

In this work, we propose to rephrase and generalize the above definition of species scale by replacing in \eqref{oldSC} the number of species, $N_{sp}$, with a notion of entropy of species, ${\cal S}_{sp}$, to be further commented below, leading to
\begin{equation}\label{new1SC}
\Lambda_{sp} = \frac{M_{P}}{{\cal S}_{sp}^{\frac{1}{d-2}}}.
\end{equation}
Comparing with \eqref{BHentropy}, this is equivalent to stating that the entropy of $N_{sp}$ species coincides with the Bekenstein-Hawking entropy of a black hole with size given by the UV cutoff $R_{sp}\simeq \Lambda_{sp}^{-1}$, namely
\begin{equation}
{\cal S}_{sp}\simeq (R_{sp})^{d-2}M_P^{d-2}\, .\label{speciesentropy}
\end{equation}
Next, we also argue that species possess a temperature, $T_{sp}$, which is again determined by $\Lambda_{sp}$. Thus, we propose that species in quantum gravity possess thermodynamic properties governing their variation and their motion within the moduli space of the effective theory. In case in which one can identify a particular modulus field with the time coordinate, species thermodynamics eventually may become ``real", in the sense that the adiabatic time variation of species properties will be ruled by their thermodynamic properties.

We will explain how the species thermodynamics is in close analogy to black hole thermodynamics and it can indeed be derived from black hole physics, as it is already anticipated from \eqref{speciesentropy}. 
Concretely, the species entropy ${\cal S}_{sp}$ for a tower of states is given in terms of the Bekenstein-Hawking entropy of the corresponding minimal black hole of UV cutoff size. 
For particular towers, like KK towers or string towers, this minimal black hole entropy can precisely be identified with the number of species within the tower. 
It follows that the species entropy (up to small,  additive log-corrections) is just given by the number of species, namely
\begin{equation}
{\cal S}_{sp}\simeq {\cal S}_{BH,{\rm min}}  \simeq N_{sp}\, .\label{speciesentropy1}
\end{equation}

Next, the species temperature for non-BPS species can be derived from the temperature of some corresponding, non-extremal minimal black holes and it will turn out to be 
\begin{equation}
T_{sp}\simeq T_{BH,{\rm min}}\simeq \Lambda_{sp}\, .
\end{equation}
Hence the species thermodynamics also originates from a gravitational area law and the laws of species thermodynamics can be formulated in an analogous way to the black hole thermodynamics. 
The derivation of the species entropy ${\cal S}_{sp}$  via the minimal black hole entropy  indeed works well for towers of states in string theory, which are always massive in the interior of the moduli space.
For $N_0$ strictly massless or very light species, one can use an alternative perturbative computation to derive the species entropy. In this case, the species entropy contains a sizable log-correction and
is given as ${\cal S}_{sp}\simeq  N_{0}\log N_0$, which nicely reflects the statistical properties of $N_0$ massless particles.

To connect our work to previous literature, let us first mention that the original derivation of the species scale in \cite{Dvali:2007hz} was already relying on properties of black holes, while the notion of a maximal temperature of species already appeared in \cite{Dvali:2007wp}.
Moreover, the close relationship, or analogy, between species and black hole thermodynamics becomes evident in the N-portrait picture \cite{Dvali:2011aa} of black holes, namely when identifying  the black hole  entropy with the number of $N$ gravitons that are regarded as the constituents of black holes.

More recently, in the context of the swampland program \cite{Vafa:2005ui,Ooguri:2006in,Palti:2019pca}, the importance of the species scale as UV cutoff of the effective field theory  was  discussed  in \cite{Cribiori:2021gbf,Castellano:2021mmx,Lust:2022lfc,Blumenhagen:2022dbo,Cota:2022yjw,Castellano:2022bvr,vandeHeisteeg:2022btw,Cota:2022maf,Cribiori:2022nke,Castellano:2023qhp,Cribiori:2023ihv,vandeHeisteeg:2023ubh,Andriot:2023isc,vandeHeisteeg:2023uxj}.
In particular, it was emphasized that the species scale should be a function of the scalar fields and, in the context of compactifications of type II string theory on Calabi-Yau threefolds, such function is related to the topological free energy \cite{vandeHeisteeg:2022btw}, as it can be also shown from deriving the species entropy as minimal black hole entropy \cite{Cribiori:2022nke}.
Moreover, swampland distance relations for the species scale as function of the moduli were recently discussed in \cite{vandeHeisteeg:2023ubh}, while large and small black hole entropy limits and their relations to swampland distance conjectures were recently discussed in the literature
\cite{Bonnefoy:2019nzv,Cribiori:2022cho,Delgado:2022dkz}.
In our present work, we will see that the swampland bound on the species scale implies analogous bounds on the black hole and species entropies.
In the context of the cosmological constant and the (anti)-de Sitter distance conjecture (ADC) \cite{Lust:2019zwm}, we will see that imposing the Bousso bound \cite{Bousso:1999xy} to the species entropy will lead to constraints on the scaling parameter of the ADC.
Hence, as we will discuss, actually many of the distance considerations and distance bounds have their deeper origin in the species thermodynamics (or vice versa). Hence species thermodynamics will shed new light
on the swampland distance conjectures.

The paper is organized as follows. In section \ref{sec:NspandS}, we review some aspects of the species scale and present our proposal on how to relate it to the species entropy, namely by the perturbative computation of the graviton propagator, by considering the minimal black  holes and their entropy and by the topological string partition function.
(Section \ref{topf1} has overlap with work to appear \cite{Cribiori:2023sch}.) In section \ref{sec:swampbounds}, we discuss entropy bounds for the species entropy and connect it to some recent results within the swampland program.
In section \ref{Thermo}, we continue the discussion about species thermodynamics, and we derive the species energy and temperature from the minimal entropy of non-extremal black holes.
Finally, in section \ref{Thermo} we formulate the laws of species thermodynamics, which are in analogy to the laws of black hole thermodynamics. We can use them to derive general constraints on the variations of moduli in string theory, leading to a new perspective on swampland conjectures, such as the swampland distance conjecture.

\section{Number of species and species entropy}
\label{sec:NspandS}

In this section, we present several arguments supporting the conceptual generalization of \eqref{oldSC} that we are proposing, namely the replacement of the number of species $N_{sp}$ with their entropy ${\cal S}_{sp}$. 

\subsection{Three ways to derive the species scale}

The species scale is the scale at which gravity becomes strong and also non-perturbative, gravitational effects start to play an important role. Exploiting this definition, three different ways to derive the species scale have been outlined in the literature. We summarize them below.

\begin{itemize}
   
\item Consider the perturbative computation of the graviton propagator, where all light fields contribute. The species scale is the scale at which the inverse propagator becomes divergent. This immediately leads to  $\Lambda_{sp}$, as given in \eqref{oldSC}, where $N_0$ counts the number of light fields with masses much lighter than $\Lambda_{sp}$.  In addition, there are also logarithmic corrections to \eqref{oldSC}, making the connection of $\Lambda_{sp}$ to thermodynamics more evident.

\item  Consider small black holes. The entropy of the smallest possible black hole within the effective description is the species scale. In turn, this means that at $\Lambda_{sp}$ small black holes will start being produced. Thus, the species scale is clearly connected to the thermodynamics of black holes, for the entropy of the tower of species is determined by the entropy of the smallest corresponding black hole.
In other words, the species entropy ${\cal S}_{sp}$ is determined by the Bekenstein-Hawking area law, where the size of the event horizon is given by the UV cutoff length $L_{sp}\simeq 1/ \Lambda_{sp}$, such that ${\cal S}_{sp}\simeq L_{sp}^{d-2}$.

\item  Consider the derivative expansion of the effective action. The coefficient controlling the $R^2$ term is the species scale. Therefore, at the scale $\Lambda_{sp}$ higher curvature corrections become relevant. In compactfications of type II string theory on Calabi-Yau threefolds, such coefficient is given in terms of the genus-one topological string partition function \cite{vandeHeisteeg:2022btw}. The fact that it can directly be obtained from black hole arguments \cite{Cribiori:2022nke}, suggests again a relation to thermodynamics. 
Besides, (additive) log-corrections appear when making the topological string partition  modular invariant, as we will discuss in \cite{Cribiori:2023sch}.
         
\end{itemize}

\noindent
All together, these arguments support the proposed relations \eqref{speciesentropy} and \eqref{speciesentropy1} between the species scale $\Lambda_{sp}$ and the species entropy ${\cal S}_{sp}$.
In the remainder of this section, we discuss each of these arguments more in detail, presenting concrete examples.

\subsection{Perturbative derivation of the species scale}

The perturbative argument to derive the species scale \cite{Dvali:2007wp} relies on the form of one-loop quantum corrections to the graviton propagator in a theory of gravity coupled to $N$ particles. Accordingly, $\Lambda_{sp}$ is defined as the scale in which perturbation theory breaks down since the first order term is of the same order as the tree level. The one-loop propagator in a $d$-dimensional flat background is given by a certain tensor with four spacetime indices times the resummed vacuum polarization $G(p^2)$.\footnote{Notice that the propagator can be reliably calculated even if the theory is non-renormalizable, namely even if there is no way to absorb the momentum dependence into the running of the couplings.}
Neglecting unimportant factors of $2\pi$, the latter can be written schematically as \cite{Calmet:2017omb}
\begin{equation}
G^{-1}(p^2) \simeq p^2 \left(1-N_0\left(\frac{ p^2}{M_P^2}\right)^{\frac{d-2}{2}}\log\left(-\frac{p^2}{\mu^2}\right)\right) .
\end{equation}
Here, we are assuming that all species particles are massless. We denote with $N_0 = \sum_\alpha \omega_\alpha N_{\alpha}$ the weighted sum over the number of massless states with species $\alpha$, and $\mu$ is the renormalization scale, which we fix below the Planck mass $M_{P}$.
Perturbation theory breaks down at a scale $p^2 = -\Lambda_{sp}^2$, where we have
\begin{equation}
\frac{M_{P}^{d-2}}{N_0} \simeq  \Lambda_{sp}^{d-2} \log(\Lambda_{sp}/\mu).
\end{equation}
This equation is solved by the $(-1)$-branch of the Lambert $W$ function \cite{Castellano:2022bvr}
\begin{equation}\label{SpeciesScaleAndW-1}
\Lambda_{sp}^{d-2} \simeq -(d-2) \frac{M_{P}^{d-2}}{N_0} \left[ W_{-1}\left( - \frac{d-2}{N_0} \left(\frac{M_P}{\mu}\right)^{d-2} \right) \right]^{-1}
\end{equation}
and the resulting expression encodes in a closed form the one-loop quantum correction to $\Lambda_{sp}$. For large values of $N_0$, using that for  $x\to 0^-$ one has  $W_{-1} (x) \simeq \log(-x)-\log(-\log(-x))+\dots$, we can extract the corrections as
\begin{equation}
\Lambda_{sp}^{d-2} \simeq \frac{M_{P}^{d-2}}{N_0 \log N_0}. 
\end{equation}
Using Stirling's formula, we can then rewrite this in the form (again for large $N_0$)
\begin{equation}
\label{speciesEntropy}
    \Lambda_{sp}^{d-2} \simeq \frac{M_{P}^{d-2}}{ \log N_0!} = \frac{M_{P}^{d-2}}{ \log\Gamma(N_0-1)}\,.
\end{equation}
Indeed, the denominator is reminiscent of the entropy of a micro-canonical ensemble of $N_0$ indistinguishable particles. 

In short, we have seen that including one-loop quantum corrections of $N_0$ massless particles effectively amounts to replace $\log (e^{N_0}) \to\log N_0!$ into the original definition \eqref{oldSC} of species scale. This paves the way for 
the proposed conceptual step: the number of species should rather be understood as an entropy. 
Therefore we define
the species entropy for $N_0$ massless species as
\begin{equation}
\label{speciesEntropyn0}
  {\cal S}_{sp}\simeq  N_0 \log N_0 \sim \log N_0!.
\end{equation}

Crucially, note that for a tower of states, like KK modes in string compactification, this multiplicative log-correction is absent, since the sum goes over massive states with masses up to the UV cutoff $\Lambda_{sp}$.  Indeed, let us consider the case of a single compact direction with radius $R$. A standard heat-Kernel calculation with Schwinger parameter $t$ gives
\begin{equation}
N_{sp} \simeq \int_0^\infty \frac{dt}{t} e^{-t \mathcal{M}^2} \simeq -\frac{1}{2} {\rm Tr}\log \mathcal{M}^2,
\end{equation}
where $\mathcal{M}=n/R$ is the KK spectrum operator with level $n$. Regularizing the trace up to a number $N_{KK}$ of states and then taking this to be large, we have
\begin{equation}
N_{sp} \simeq  -\lim_{N_{KK} \rightarrow \infty} \sum_{n \neq 0}^{N_{KK}} \log \frac{n^2}{R^2} = \lim_{N_{KK} \rightarrow \infty} \log{\frac{N_{KK}!}{R^{N_{KK}}}}
\end{equation}
and this gives the desired result, $N_{sp}\simeq N_{KK}$, if one uses that $N_{KK}\simeq R$, which is formula \eqref{vol}, and the fact that $n^n/n!\sim e^n$.
In any case, the thermodynamic nature of the entropy of a tower of states can be seen most directly with the second argument to derive the species scale, which we will now discuss.

\subsection{Minimal Black hole derivation of the species scale}\label{minimalBH1}

The non-perturbative argument calculates the species scale as the scale set by the smallest possible black hole which can be described reliably within the effective theory.
In quantum gravity, one would naively assume that the smallest possible black hole is of Planck size, i.e.~$R_{BH}\simeq 1/M_P$. However, in the presence of species there is a bound on the Schwarzschild radius $R_{BH}$ of any black hole in the effective theory, namely
\begin{equation}
    R_{BH}\geq L_{sp}={\frac{1}{\Lambda_{sp}}}, \qquad \implies \qquad 
    {\cal S}_{BH} \geq {\cal S}_{sp}.
\end{equation}
Therefore, we define the minimal black hole entropy to be given by 
\begin{equation}
{\cal S}_{BH,min} :={\cal S}_{sp} = \Lambda^{2-d}_{sp}M_P^{d-2} \,.
\end{equation}
In this way, the species entropy of (a tower of) states with masses below  $\Lambda_{sp}$ corresponds to the entropy of a black hole with radius of the order of the UV cutoff length $L_{sp}=\Lambda^{-1}_{sp}$.
As we will discuss later in section \ref{LambdaBousso} for the case of a KK tower of states, the species entropy, being the Bekenstein-Hawking entropy of a region of spacetime of radius $L_{sp}$, is restricted by the Bousso bound.

In the following, we provide an algorithm to determine the size of the smallest possible black hole that can be described within effective theory. Since we will work mainly with examples from string theory, this algorithm will be applicable to charged black holes and it will be built on recent works \cite{Bonnefoy:2019nzv, Cribiori:2022cho,Delgado:2022dkz,Cribiori:2022nke}. An alternative strategy, valid for Schwarzschild black holes, will be provided in \cite{toappear} in the context of the N-portrait picture \cite{Dvali:2011aa}, where the smallest possible black hole will have a number of constituents $N$ such that $N=\mathcal{S}_{sp}$.
Focusing on charged black holes, we propose to determine the smallest possible one by eliminating all charge dependence from the entropy in a precise manner, namely by setting the magnetic (or electric) charges to a minimal value while at the same time requiring that the entropy grows in the weak coupling or large volume limit. This procedure works thanks to the combination of two effects. On the one hand, when setting charges to a minimal value the black hole horizon shrinks up to a point in which corrections to the spacetime effective action can lead to a modification of the geometry. Thus, this operation produces a small black hole but can potentially invalidate the effective description. To disentangle this latter undesired effect, we require that the expression for the minimal entropy should be a function of the moduli which grows in the weak coupling or large volume limit. In this way, we get a small black hole in a regime in which the effective description is trustworthy.

As we are now going to show in explicit examples of charged black holes, the proposed algorithm provides the expected result for the species scale. Let us consider extremal dilatonic black holes coming from heterotic string theory compactified on $K3\times T^2$. Their entropy is given by \cite{Ferrara:1996dd}
\begin{equation}
{\cal S}_{BH} = \pi pq=\pi p^2g_s^{-2}\, ,
\end{equation}
where $p$ and $q$ are magnetic and electric charges with the same sign and $g_s=e^\phi=\sqrt{p/q}$ is the string coupling, with $\phi$ being the dilaton field.
The minimal non-vanishing entropy is obtained for a charge configuration with $p=1$ and thus one finds
\begin{equation}
{\cal S}_{sp}={\cal S}_{BH,min}\simeq g_{s}^{-2} \,,
\end{equation}
in accordance with equation (\ref{gs}) for species given by a tower of string states. Notice that by choosing to eliminate the magnetic charge $p$ instead of the electric charge $q$, and setting then $q=1$, we would get $\mathcal{S}_{BH, min} \simeq g_s^2$ which does not grow in the weak coupling limit and thus cannot be used to determine the species scale according to our algorithm.\footnote{More precisely, this would be the smallest possible black hole in the S-dual description.}

Let us consider a second example, namely BPS black holes arising from type IIA compactifications on a Calabi-Yau threefold. These black holes can be constructed as superpositions of three kind of D4-branes, wrapped around $p^i$ Calabi-Yau
four-cycles and $q$ D0-branes. Being extremal, their temperature is vanishing, but we can still use them to read off the minimal black hole entropy, i.e.~the corresponding species entropy.
In section \ref{Thermo} we will look at their associated non-extremal minimal black holes, which will provide us with the corresponding species temperature.

In the large volume limit, the prepotential of low energy effective supergravity action is dominated by the classical term $F =-\frac16 C_{ijk}X^i X^j X^k/X^0$, with $C_{ijk}$ the triple intersection numbers of the Calabi-Yau. For a black hole solution supported by positive non-vanishing charges $-q^0\equiv q$ and $p^i$, the attractor mechanism \cite{Ferrara:1996dd} fixes the entropy to be \cite{Behrndt:1996jn}
\begin{equation}
{\cal S}_{BH} = 2\pi \sqrt{\frac q6 C_{ijk}p^i p^j p^k}.
\end{equation}
On the other hand, the total volume of the Calabi-Yau at the black hole horizon is given by $\mathcal{V}_H=\sqrt{\frac{q^3}{\frac16 C_{ijk}p^ip^jp^k}}$. The minimal non-vanishing entropy is obtained for a charge configuration with $\frac16 C_{ijk}p^ip^jp^k=1$ and thus one finds
\begin{equation}\label{CYVolumeBH}
{\cal S}_{sp}={\cal S}_{BH,min} \simeq \sqrt{q} = \mathcal{V}_H^\frac{1}{3} \simeq N_{KK} .
\end{equation}
As for the previous example, by eliminating the magnetic charges instead of the electric one, we would have eventually obtained an entropy which does not grow at large volume and thus cannot be used in our algorithm. 
We see that the species scale is governed by the third root of the total volume of the Calabi-Yau threefold. Effectively, $\mathcal{V}_H^\frac13\simeq R^2$ is the volume of a two cycle of radius $R$, which is the smallest possible cycle on a simply connected Calabi-Yau.

This result is in accordance with \eqref{vol} for species given by a tower of KK states and it confirms that the species entropy is given in terms of the number KK modes.
Hence, in this spirit, for a compactification from $D=d+n$ dimensions down to $d$ space-time dimensions on an isotropic compact $n$-dimensional space of radius $R$, with number of KK particles in the tower below $\Lambda_{sp}$ given by
\begin{equation} N_{sp}=(\Lambda_{sp}R)^n\, ,
\end{equation}
and with the corresponding species scale 
\begin{equation} \Lambda_{sp}=\frac{(M_P)^{\frac{d-2}{ D-2}}}{ R^{\frac{n}{D-2}}},\label{KKspecies}
\end{equation}
 the associated entropy of the KK tower is given as
\begin{equation} {\cal S}_{sp}=(M_PR)^{\frac{(d-2)n }{D-2}}\, .\label{KKentropy}
\end{equation}
Note that this KK tower entropy does not receive multiplicative log-corrections of the form $\log N_{KK}$, but only a small additive correction proportional to $\log N_{KK}$, as also discussed in an example in the next subsection.
This is because in order to derive $\Lambda_{sp}$ one is summing over a massive tower of KK states, in contrast to the case of summing over $N_0$ light or massless (indistinguishable) species, where we obtained the multiplicative log-corrections $\log N_0$ from the massless loop calculation.
One may view this result as the fact that the $N_{KK}$ KK species contribute to the species entropy in the same way as $N_0$ massless particles obeying the relation that $N_0\log N_0=N_{KK}$, and hence always $N_0<N_{KK}$.

\subsection{Species entropy and the topological string free energy}\label{topf1}

The number of species should be understood as an entropy also when looking at black holes with higher derivative corrections.
In general, higher derivative corrections render the use of the Bekenstein-Hawking formula for the entropy of black holes invalid and one has rather to employ the Wald formula \cite{Wald:1993nt}. 
As is know from \cite{Antoniadis:1993ze}, certain higher-derivative terms in low-energy effective supergravity actions can be calculated via topological string computation.
Then, the conjecture of \cite{Ooguri:2004zv} relates the topological amplitude to the microscopic degeneracies of the black hole.

The formula obtained by Wald calculates the entropy in terms of the full lagrangian $\mathcal{L}$ including higher derivative corrections as \cite{Wald:1993nt}
\begin{eqnarray}
    {\cal S}_{BHW} = 2\pi \int_\Sigma  d\Omega \sqrt{h} \frac{\partial\mathcal{L(R)}}{\partial R_{\mu\nu\rho\sigma}} \epsilon^{\mu\nu} \epsilon^{\rho \sigma} ,
\end{eqnarray}
where $h$ is the induced metric on the horizon $\Sigma$, $\epsilon^{\mu\nu}$ the binormal with indices $\mu,\nu=0,1$ in our conventions. Let us consider BPS black holes in $\mathcal{N} = 2$ supergravity in four dimensions. Then, for a generic prepotential $F=F(X,A)$, where $X^\Lambda$ are vector multiplets and $A$ the chiral graviphoton background coupling to the higher derivative corrections, the entropy is given by \cite{LopesCardoso:1998tkj}
\begin{equation}
\mathcal{S}_{BH} =\pi\left[Z\bar Z +4{\rm Im}(A\partial_A F(X,A))\right], 
 \end{equation}
with $ Z\bar Z =q_\Lambda X^\Lambda - p^\Lambda \partial_\Lambda F(X,A)$, and it has to be evaluated at the attractor point
\begin{equation}
p^\Lambda = -2{\rm Im} X^\Lambda, \qquad q_\Lambda = -2 {\rm Im} \partial_\Lambda F(X,A),\qquad A=-64 .
\end{equation}
As noticed in \cite{Ooguri:2004zv}, one can exploit the homogeneity relation $X^\Lambda \partial_\Lambda F+2A \partial_A F=2F$ together with the attractor equations to rewrite the entropy as 
\begin{equation}
\label{SBHLegtransf}
\mathcal{S}_{BH} =\mathcal{F}(\phi,p)-\phi^\Lambda \frac{\partial \mathcal{F}}{\partial \phi^\Lambda}. 
\end{equation}
Here, we introduced a real part $\phi^\Lambda/(2\pi)$ of $X^\Lambda$ and the function $\mathcal{F}(\phi,p)=4\pi{\rm Im}F$, such that $\frac{\partial \mathcal{F}}{\partial \phi^\Lambda} = -q_\Lambda$. The expression \eqref{SBHLegtransf} recast the entropy as the Legendre transform of $\mathcal{F}$. The latter can be understood as a (mixed) black hole free energy, in the sense that the (mixed) black hole partition function is given by $\mathcal{Z}_{BH} = e^\mathcal{F}$.
At this point, the number of species can be found by minimizing the entropy \eqref{SBHLegtransf}. Since at the minimum axions are set to zero, $\phi^\Lambda=0$, we get
\begin{equation}
\label{N2BHEntropy}
N_{sp} = {\rm min}\left[\mathcal{F}(\phi,p) + \phi^\Lambda q_\Lambda\right] = \mathcal{F}_* \equiv \mathcal{S}_{sp},
\end{equation}
where a star indicates the attractor point. Thus, we can write the species scale as 
\begin{equation}
\Lambda_{sp} = \frac{M_P}{\sqrt{\mathcal{F}}}
\end{equation}
This argument was presented in \cite{Cribiori:2022nke} and reproduces the result of \cite{vandeHeisteeg:2022btw}.

As an illustrative example, let us take as target space of the topological string a torus $T^2$. For this setting, the genus one free energy of the topological string is \cite{Bershadsky:1993ta} 
\begin{eqnarray}
    F_1 = -\log{\tau_2 |\eta(\tau)|^4}
\end{eqnarray}
where $\tau = \tau_1 + i\tau_2$ is the complexified  K\"ahler parameter, and $\eta(\tau)$ is the Dedekind eta function. 
In the decompactification limit $R^2 \equiv \tau_2 \xrightarrow[]{} \infty$  we expect a large number of KK species and we obtain
that
\begin{eqnarray}
    F_1 = -\log{[R^2 |\eta(iR)|^4]} \simeq  \frac{2\pi R^2}{3} -2\log{R} + 4  e^{-2\pi R^2} \, .
\end{eqnarray}
According to our proposal, we finally get
\begin{eqnarray}
    {\cal S}_{sp} =F_1\simeq \frac{2\pi R^2}{3}\sim N_{KK} \, .
\end{eqnarray}
We also see that the log-corrections in terms of $R$ are additive and negligible in large $R$ limit.

\vspace{.7cm}

To summarize, we have seen in the last three subsections that  $N_{sp}$ should rather be understood, or better replaced, by the entropy of species ${\cal S}_{sp}$. This suggest that one can understand the tower of species with masses up to 
$\Lambda_{sp}$ using the language of thermodynamics. In particular, the entropy of the tower is responsible for lowering the quantum gravity cutoff from $M_P$ to $\Lambda_{sp}$.

Before proceeding, one comment is in order:
let us notice that since we are always considering a large value of ${\cal S}_{sp}$, in our discussion the (gran-)canonical ensemble reduces to the microcanonical. 
Indeed, in the thermodynamic limit, here expressed by the large $S_{sp}$ limit, the canonical and gran-canonical observables can be calculated in the micro-canonical ensemble. 
In fact, without supposing any interaction between particles (at this approximation), we obtain the logarithm of the quantum number of species micro states:
\begin{equation}
\left(\frac{\Lambda_{sp}}{M_P}\right)^{d-2} \sim \frac{1}{\log n_{\rm micro}}\sim \frac{1}{\log e^{S_{sp}}},
\end{equation}
up to a constant.

\section{Entropy bounds and the Swampland}
\label{sec:swampbounds}

In this section, we connect the species entropy introduced and motivated before to various swampland ideas and conjectures, such as the distance conjecture, the Dark Dimension scenario and the de Sitter conjecture.

\subsection{Entropy bound and the Swampland Distance conjecture}

The Swampland Distance Conjecture postulates that a tower of light particles emerges when we approach to the boundary of the moduli space. 
The appearance of these new degrees of freedom will 
certainly produce entropy variation which decrease the quantum gravity cutoff $\Lambda_{sp}$ \eqref{new1SC}.
Recently, in \cite{vandeHeisteeg:2023ubh} a new bound on the variation of $\Lambda_{sp}$ as a function of the moduli fields was derived and argued to be valid also in the interior of the moduli space. Such a bound reads  
\begin{eqnarray}\label{SpeciesScaleBound}
    \bigg|  \frac{\Lambda_{sp}^{'}(\phi)}{\Lambda_{sp}(\phi)} \bigg|^2\lesssim \frac{c_{sp}^2}{M_p^{d-2}},
\end{eqnarray}
for $c \sim \mathcal{O}(1)$ constant and where primes on functions denote partial derivatives. 

Our goal is to understand how the species entropy change in the moduli space and how it could be connected to other Swampland conjectures. 
This can be done by recasting the species scale in terms of the entropy \eqref{new1SC}. Then, exploiting the dependence of our thermodynamic variables in terms of the moduli fields, we obtain
\begin{equation}
    \bigg|  \frac{{\cal S}_{sp}^{'}(\phi)}{{\cal S}_{sp}(\phi)} \bigg|^2 \lesssim \frac{(d-2)^2 c_{sp}^2}{M_p^{d-2}}.
\end{equation}
In the positive region of the derivative, we can integrate this inequality and write an upper bound on the species entropy (in Planck units)
\begin{eqnarray}\label{SpeciesEntropyDC}
    {\cal S}_{sp}(\phi) \lesssim {\cal S}_{sp}(\phi_0) e^{(d-2)c_{sp}\Delta \phi}, 
\end{eqnarray}
where the asymptotic behavior is satisfied as we approach the boundary of moduli space.  There, we get a weakly coupled light tower of states, whose mass scales according to the swampland distance conjecture as
\begin{equation}
    m(\phi) \sim  e^{-c_{DC} \phi},
\end{equation} 
in Planck units and where $c_{DC} \sim \mathcal{O}(1)$ constant. Via $\Delta\phi$, we can connect the above two relations and obtain a Species Distance Conjecture 
\begin{eqnarray}\label{SpeciesEntropyDistance}
    {\cal S}_{sp} (\phi)\lesssim m(\phi)^{-\frac{(d-2)c_{sp}}{c_{DC}}} \equiv m(\phi)^{-\gamma}\,,
\end{eqnarray}
for the constant parameter $\gamma = \frac{(d-2)c_{sp}}{c_{DC}}$
This is indeed the Black Hole Entropy Distance Conjecture (BHEDC) \cite{Bonnefoy:2019nzv}, now expressed in terms of the species entropy. 
From this calculation we can see that the Swampland Distance conjecture and \eqref{SpeciesScaleBound} imply the Species Entropy Distance \eqref{SpeciesEntropyDistance}, and also to the BHEDC \cite{Bonnefoy:2019nzv}. 
The limit of large black hole entropy as well as of large species entropy is at infinite distance in the parameters space and, in either of the two cases, the associated states in the tower become light.
Actually, as argued in \cite{Bonnefoy:2019nzv}, for large entropy the infinite tower of light modes of the BHEDC may describe the degrees of freedom for the transitions between the black hole micro states. Indeed, in the large black hole entropy limit, the frequency differences between micro states of a black hole approach zero, composing a complete degenerate set of micro states. 
The same behavior holds true for the differences in the tower masses, when one approaches the large species entropy limit. 
Therefore, we conclude that the Swampland Distance Conjecture may be seen as a Species (or Entropy) Distance Conjecture.

\subsection{Cosmological constant, Bousso bound and the Dark Dimension}\label{LambdaBousso}

In this section, we compare the cosmological constant $\Lambda_{cc}$ with the species scale.
The measured value of the cosmological constant at present is \cite{Planck:2018vyg}
\begin{equation}
\Lambda_{cc}\simeq 10^{-122}M_P^4\, .
\end{equation}
As explained in the previous sections, the species scale $\Lambda_{sp}$ provides the UV quantum gravity cutoff of the theory, being related to the species entropy $S_{sp}$, which is the 
Bekenstein-Hawking entropy of a region in space-time of length scale 
$L_{sp}\simeq\Lambda_{sp}^{-1}$. On the other hand, on a positive energy density background the IR cutoff length of the theory is provided by the Hubble radius $L_{cc}$, i.e.~the curvature radius, related to cosmological constant as
\begin{equation}
L_{cc}^2\simeq \Lambda_{cc}^{-1}M_P^2\, .
\end{equation}
One can also associate an entropy to the Hubble radius, which is the  Gibbons-Hawking entropy ${\cal S}_{GH}$ \cite{Gibbons:1977mu}, obeying an area law of the form
\begin{equation}
{\cal S}_{GH}\simeq L_{cc}^{d-2}M_P^{d-2}\simeq \Lambda_{cc}^{-\frac{{d-2}}{2}}M_P^{2(d-2)}\, .
\end{equation}
With the value of $\Lambda_{cc}$ measured at present, the Gibbons-Hawking entropy is of order ${\cal S}_{GH}\simeq10^{122}$.

Besides, we can apply the Bousso bound \cite{Bousso:1999xy}, requiring that the species entropy of a certain subregion in spacetime is always less or equal than the Gibbons-Hawking entropy (see also the discussion in \cite{Ooguri:2018wrx}), namely
\begin{equation}\label{SpeciesScaleBousso}
{\cal S}_{sp}\leq {\cal S}_{GH}\quad \Longleftrightarrow  \quad {L}_{sp}\leq {L}_{cc} \,,
\end{equation}
implying that the UV cutoff $L_{sp}$ is always a lower bound of the IR cutoff $L_{cc}$.\footnote{As noticed by \cite{Castellano:2021mmx}, through \eqref{SpeciesScaleBousso} it is possible to obtain an inequality between the cosmological constant and the tower scale \eqref{KKentropy} as $m \gtrsim \Lambda_{cc}^\alpha$ 
where $\alpha$ is an exponent which depends on the dimensions of the spacetime, and on the degeneracies of states. This relation, which comes from entropy argument, becomes indeed the Anti-de Sitter Distance conjecture \cite{Lust:2019zwm} in the large $N_{sp}$ limit, or small $\Lambda_{cc}$.} 
We assume the existence of a mixing between the UV cutoff and the IR cutoff, which can be stated requiring that 
$L_{cc}$ depends on the number of species, i.e. $L_{cc}^{d-2}\simeq N_{sp}^\epsilon M_P^{2-d}$, for a certain parameter $\epsilon$.
Then, the Bousso bound simply becomes
\begin{equation}
\epsilon \geq 1\, .
\end{equation}

The precise determination of the parameter $\epsilon$ can be provided by using the Anti-de Sitter conjecture (ADC) \cite{Lust:2019zwm}, here applied to the case of a positive cosmological constant.
Concretely, the ADC states that in the limit of vanishing cosmological constant there must be a light tower of states, whose mass scale $m$ is parametrically related to $\Lambda_{cc}$ in the following way
\begin{equation}
\text{ADC}:\qquad m\simeq \Lambda_{cc}^\alpha M_P^{1-4\alpha}\,,
\end{equation}
where the parameter $\alpha$ is bounded as 
\begin{equation}
{\frac{1}{ d}}\leq \alpha\leq {\frac{1}{2}}\,.
\end{equation}
In particular the upper bound is model independent and originates from the Higuchi bound \cite{Higuchi:1986py}, while the lower bound can be argued for by considering 1-loop string potentials in $d$ dimensions \cite{Montero:2022prj}.

In the following, we apply the Bousso bound to a KK tower originating from compactification from $D=d+n$ dimensions down to $d$ space-time dimensions on an isotropic compact $n$-dimensional space of radius $R$.
In this setup, the tower mass scale is given as $m=1/R$. 
The species scale and the associated entropy are given in \eqref{KKspecies} and \eqref{KKentropy}.
Due to UV/IR mixing, the species entropy and the Gibbons-Hawking entropy are non independent from each other, and their ratio is given in terms of $R$ as
\begin{equation}
\frac{{\cal S}_{GH}}{{\cal S}_{sp}}=(RM_P)^\frac{(D-2-2n\alpha)(d-2)}{(D-2)2\alpha}\, .
\end{equation}
Furthermore, the Bousso bound now reads
\begin{equation}
(RM_P)^{{\frac{(d-2)n}{ D-2}}}\leq (RM_P)^{\frac{d-2}{2\alpha}}\, .
\end{equation}
The parameter $\epsilon$ can now easily be computed and it is given by 
\begin{equation}
\epsilon={\frac{D-2}{2n\alpha}}\, .
\end{equation}
Therefore, the Bousso bound $\epsilon \geq 1$ can be translated into the following condition on the space-time dimension $d$ in terms of the number of compact dimensions $n$:
\begin{equation}
d\geq n(2\alpha-1)+2\, .\label{dbound}
\end{equation}
For  $\alpha=1/2$, which is in fact model independent and follows from the Higuchi bound,
we simply obtain as condition from the Bousso bound that $d\geq 2$. For all other values of $\alpha<1/2$, the bound becomes weaker.

Turning the logic around, the same Bousso bound $\epsilon \geq 1$ can be translated into the following condition on the parameter $\alpha$:
\begin{equation}
\alpha\leq {\frac12}+{\frac{d-2}{2n}}\, .\label{alphabound}
\end{equation}
If we require this relation to be true for all possible values $d,n$ and in particular also for $d=2$, it then follows that 
$\alpha\leq {1/2}$. This is just the bound on $\alpha$, which was previously obtained from the Higuchi bound, namely from the requirement of the unitarity of the effective field theory with spin-two KK gravitons.
So we see that the Bousso bound and the Higuchi bound lead to equivalent results.

Let us finally consider the concrete case of the Dark Dimension scenario \cite{Montero:2022prj} and its implications
\cite{Anchordoqui:2022ejw,Anchordoqui:2022txe,Blumenhagen:2022zzw,Gonzalo:2022jac,Anchordoqui:2022tgp,Anchordoqui:2022svl,Anchordoqui:2023oqm}. This scenario corresponds to $d=4$, $n=1$ and $\alpha=1/4$ and it expresses the cosmological constant in terms of one large extra dark dimension of size $R_{dd}\simeq \Lambda^{-1/4}\simeq 1\mu{\rm m}$
(here we neglect the constant $\lambda\simeq 10^{-3}$ which also enters this relation). The tower entropy of the dark KK gravitons,
\begin{equation}
{\cal S}_{dd}\simeq R_{dd}^{2/3}M_P^{2/3}\simeq 10^{20}\,,
\end{equation}
corresponds  to a species scale $\Lambda_{dd}\simeq R^{-1/3}M_P^{2/3}\simeq 10^{-10}M_P$. 
This entropy is indeed very tiny compared the Gibbons-Hawking entropy ${\cal S}_{GH}\simeq 10^{122}$ and the Bousso bound is clearly satisfied.

\section{Thermodynamic species energy and temperature}\label{Thermo}

In the previous sections, we introduced and motivated the notion of species entropy and related it to the more familiar concept of black hole entropy and to various swampland conjecture.
Now, we would like to take a step further and identify other thermodynamic quantities associated to the species, such as temperature, energy  and the heat capacity.
 
To start, we recall some well known relations from the $d$-dimensional Schwarzschild black hole solution. Its mass $M_{BH}$, temperature $T_{BH}$ and entropy ${\cal S}_{BH}$ are given in terms of the black radius $R_{BH}$ as follows
(in Planck units and omitting numerical prefactors):
\begin{eqnarray}
M_{BH}&=&\bigl(R_{BH}\bigr)^{d-3}\, ,\nonumber\\
T_{BH}&=&\bigl(R_{BH}\bigr)^{-1}\, ,\nonumber\\
{\cal S}_{BH}&=&\bigl(R_{BH}\bigr)^{d-2}\, .
\end{eqnarray}
From that we derive the universal relation 
\begin{equation}
{\cal S}_{BH}T_{BH}^{d-2}=M_P^{d-2}\, ,\label{STrel}
\end{equation}
which will be suggestive in what follows.

Let us now translate these quantities to the language of species thermodynamics. 
We will consider some generic non-BPS species that can interact among each other and can also decay.
First, observe that the relations \eqref{STrel} and \eqref{new1SC} look identical.
Hence, the relation \eqref{STrel} may be rewritten, defining the species temperature as
\begin{equation} \label{SpeciesTemp}
    T_{sp} := \frac{M_P}{{\cal S}_{sp}^{\frac{1}{d-2}}} \equiv \Lambda_s\, .
\end{equation}

This relation can be verified by using the argument of minimal black hole entropy, but now applied to the case of non-extremal black holes with temperature (extremal black holes have vanishing temperature).
These small, non-extremal black holes in four dimensions can be constructed as follows. We start with the extremal, charged ${\cal N}=2$ black hole, which we considered in section  \ref{minimalBH1}.
Its minimal black hole entropy in terms of the internal KK volume was given in \eqref{CYVolumeBH}. As recently studied e.g.~in \cite{Cribiori:2022cho}, we can then construct associated non-extremal black hole solutions with the same
electric and magnetic charges, but now with non-vanishing temperature.
Specifically we can obtain them  by turning on the non-extremality parameter $c_0={\cal S}T$. This step can be done in such a way that the internal Calabi-Yau volume ${\cal V}_H$ stays the same as in the extremal case. Furthermore, we here set 
${\cal V}_H$ equal to the Calabi-Yau volume at spatial infinity, i.e.~${\cal V}_H={\cal V}_\infty\equiv{\cal V}$. Then, one gets the following relation between entropy and temperature (see eq.~(4.44) in \cite{Cribiori:2022cho}, up to constant factors):
\begin{equation}
{\cal S}_{BH}^3\simeq\bigl(\sqrt {{\cal V}^{1/3}p+{\cal S}_{BH}^2T_{BH}^2}+{\cal S}_{BH}T_{BH}\bigr)^6\, .
\end{equation}
Here we have set all magnetic charges equal to each other, i.e. $p_i=p$. For the minimal black hole we again set $p=1$ and we make the identification ${\cal S}_{BH,{\rm min}}\equiv{\cal S}_{sp}$ and $T_{BH,{\rm min}}\equiv T_{sp}$,
which leads to the following relation:
\begin{equation}
{\cal S}_{sp}^3\simeq\bigl(\sqrt {{\cal V}^{1/3}+{\cal S}_{sp}^2T_{sp}^2}+{\cal S}_{sp}T_{sp}\bigr)^6\, .
\end{equation}
Taking the large entropy limit ${\cal S}_{sp}\rightarrow\infty$, the relation (\ref{SpeciesTemp}) is indeed obtained for $d=4$.

Let us provide an alternative argument and derive the expression \eqref{SpeciesTemp} for the species temperature
using the first law of thermodynamics.  
In the case of simply thermal reversible processes, or analogously in the case of an adiabatic motion of the moduli fields,  we can write
\begin{equation}\label{EnergyEntropyRel}
     dE =  {T d{\cal S}} + \dots,
\end{equation}
with
\begin{equation}\label{tempRel}
\frac{1}{T} = \dfrac{\partial {\cal S}}{\partial E} \, .
\end{equation}
In order to calculate \eqref{tempRel}, we need to understand how the species energy is defined. A meaningful definition should involve energy quantities varying over the moduli space of the theory, i.e.~the species scale itself in this case. 
Thus, we postulate
\begin{equation}
    E_{sp} \simeq \Lambda_s^a,
\end{equation}
in Planck units and omitting numerical factor for now. It is easy to check that for positive power $a$, the temperature \eqref{tempRel} is always a negative defined quantity. Hence, the positivity constraint requires  $a < 0$.
Using again the similarity with black hole physics, we identify $E_{sp}$ with $M_{BH}$. This immediately leads to the following relation for the species energy (we keep working in Planck units unless otherwise specified)
\begin{equation}
    E_{sp} \simeq \Lambda_{sp}^{3-d}=\bigl({\cal S}_{sp}\bigr)^{{\frac{d-3}{d-2}}}\, ,\label{energy}
\end{equation}
i.e. $a =3-d$, which is indeed always negative in four or higher space-time dimensions.\footnote{Although we are not keeping track of constant coefficients, the correct factors are $E_{sp} = b/\Lambda_s, \; T = b \Lambda_s/2$ with $b = 1/2\pi$ if one wants the same factor of black hole thermodynamics.}

To verify this relation in a more general set-up, let us compute the species energy of a tower of states. For concreteness, we use the well known formula \eqref{oldSC}, and, in this approximation, we consider $N_{sp}$ species coming from a KK tower associated to a single compact dimension. In this simple case, the finite energy difference between two species is given by $\Delta E \simeq \Lambda_{sp}/N_{sp} \sim {\cal S}_{sp}^{(1-d)/(d-2)}$ (in units of $M_P$), and we can simply describe the levels by 
$E_k = k~\Delta E$ with
$k=1,\dots , N_{sp}$.
Finally, we can sum over all tower  energies and write the total energy $E_{sp}$ of the tower as
\begin{equation} \label{EnergyRel}
E_{sp} = \sum_{k=1}^{N_{sp}} E_k \simeq  \Lambda_{sp}^{3-d}=\bigl({\cal S}_{sp}\bigr)^{{\frac{d-3}{d-2}}}\,.
  \end{equation}
This agrees with \eqref{energy} and it indicates that the species energy is just the total energy of all particles in the KK tower. A more general proof, also including a tower of string states is presented in appendix \ref{AppendixA}.

Next, using \eqref{EnergyRel} as definition of energy we can calculate the species temperature \eqref{tempRel} and find the result \eqref{SpeciesTemp} discussed above, with the correct coefficients. Hence, we can obtain the same black hole results via species tower consideration.

Thanks to this choice of the species energy, we are able to calculate other thermodynamic quantities. As for Schwarzschild black holes, the chemical potential, i.e.~the energy that can be absorbed or released due to a change in the number of particles (keeping the entropy fixed) is zero. 
For Reissner–Nordstr\"om black holes, instead, a new independent variable is available, namely the charge $Q$ and its conjugate variable $\Phi$. In this case, a term $\Phi dQ$ must be added to \eqref{EnergyEntropyRel}, and the relation between energy and entropy changes.

Furthermore, we can compute the species heat capacity as
\begin{equation}
   C_{sp} = \frac{dE_{sp}}{dT_{sp}} = - \frac{d-3}{T_{sp}^{d-2}}.
\end{equation}
It tells that the heat capacity is always negative, which means that the system loses energy radiating particles spontaneously. Due to this process, the number of species decreases and the species scale increases, i.e.~the system spontaneously evolves towards the bulk of the moduli space via quantum effects. This result indeed highlights the Hawking radiation in a tower of species. We hope to come back to this in the future.

As a final cross check, we look back at section \ref{topf1}. Indeed, we now have at our disposal all of the quantities needed to understand if the species free energy, defined as $F_{sp}= E_{sp}-T_{sp}\mathcal{S}_{sp}$, coincides with the topological string free energy.  
Using the species temperature $T_{sp} \equiv \beta^{-1}_{sp} = 1/\sqrt{\mathcal{S}_{sp}}$ and expressing the topological free energy in \eqref{N2BHEntropy} as a thermodynamic quantity, $\mathcal{F} = - \log{\mathcal{Z}_{\textit{top}}} \equiv \beta F_{top}$, we can identify the temperature $\beta^{-1}$ with the species temperature $\beta_{sp}^{-1}$, obtaining
\begin{eqnarray}
    {\cal S}_{sp} \sim \mathcal{F} = \sqrt{{\cal S}_{sp}} F_{top} \implies F_{top} \simeq \sqrt{{\cal S}_{sp}} \simeq F_{sp}.
\end{eqnarray}
This indicates that the topological string free energy $F_{top}$ can be seen approximately as the species free energy $F_{sp}$.

\section{The laws of species thermodynamics}
\label{sec:thermolaws}

In the previous section, we identified thermodynamic quantities associated to a tower of species, and we noticed they are similar to those describing black hole thermodynamics. 
In fact, after realizing that gravitational (or species) energy variation is due variation of the entropy ${\cal S}_{sp}$, we have been able to write the first law of thermodynamics for the species.
Now, thanks to the identification of the temperature and the energy of the system, we can propose the four laws of species thermodynamics. As we will show, they will automatically include some primary swampland constraints.

Since the standard laws of thermodynamics involve time and the time variations of entropy and temperature, let us first specify what quantity plays the role of time in the context of species. Following the swampland approach, where one considers the adiabatic motion within the scalar moduli space towards its boundary at infinite distance, we postulate that the time is given by a normalized scalar fields $\phi$, like the dilaton field or the overall volume modulus, which is adiabatically moving along a geodesic line in moduli space.
For time-dependent string backgrounds, it is then possible to identify the scalar field with the real time coordinate, i.e.~$\phi\equiv t$, as it is the case e.g.~for the linear dilaton background.

With this in mind, we are ready to state the four laws of species thermodynamics.

\begin{itemize}
    \item {\bf Zeroth law of species thermodynamics.}

If the temperature $T_{sp}(\phi)$ at any two points of the moduli space coincides with that of a third point, then these two points have the same temperature. Namely $\forall \phi_0, \phi_1, \phi_2 \in \mathcal{M}$ such that $ T_{sp}(\phi_1)= T_{sp}(\phi_0)$ and $T_{sp}(\phi_2)= T_{sp}(\phi_0)$, then $T_{sp}(\phi_1) = T_{sp} (\phi_2)$. In the case in which the analogy with the Schwarzschild black hole holds, as assumed in the previous section, one can state that the temperature of the moduli space is constant along orbits with the same quantum gravity cutoff $\Lambda_{sp}(\phi)$.

    \item {\bf First law of species thermodynamics.}

    Any two neighboring stationary towers are related by
 \begin{equation}
        \delta E_{sp} = \, T_{sp} \, \delta {\cal S}_{sp} + \Phi \delta Q + \dots
    \end{equation}

  As already said, for a generic species tower the temperature $T_{sp}$ is analogous to $\frac{\Lambda_{sp}}{2\pi}$.\footnote{Since $T_{sp}\simeq{\cal S}_{sp}^{-1/2}$, the first law of thermodynamics tell us that the species energy scales as $E_{sp}\simeq {\cal S}_{sp}^{1/2}$, also verified already before.} Moreover, in order to connect it to the black holes thermodynamics we may notice again that the species scale has the same role to the surface gravity of a Schwarzschild black hole. In fact, these similarities allow us to write a dictionary between thermodynamic quantities of these two objects (we denote with $\kappa_{BH}$ the surface gravity)
\begin{align*}
   E_{sp} &\xleftrightarrow{} M_{BH}, \\
   T_{sp} \xleftarrow{} \Lambda_{sp} &\xleftrightarrow{} \kappa_{BH} \xrightarrow{} T_{BH}.
\end{align*}

\item {\bf Second law of species thermodynamics.}

The species entropy of each tower does not decrease when the system moves towards the boundary of the moduli space $\mathcal{M}$ with an adiabatic motion:
\begin{equation}
\label{2ndlaw}
\delta \Lambda_{sp}(\phi) \leq 0\, , \qquad \delta {\cal S}_{sp}(\phi) \geq 0.
\end{equation}

In other terms,
\begin{align*}
    \forall \phi_1, \phi_2 \in \mathcal{M},  &\textrm{ s.t } \min_{\phi \in \partial \mathcal{M}}\Delta(\phi_1, \phi)  >  \min_{\phi \in \partial \mathcal{M}}\Delta(\phi_2, \phi), 
    \end{align*}
    one has
    \begin{align*}
     \Lambda_{sp} (\phi_2)< \Lambda_{sp} (\phi_1)\, \qquad\Longleftrightarrow\qquad {\cal S}_{sp} (\phi_2)> {\cal S}_{sp} (\phi_1)\, ,
\end{align*}
where $\Delta(\phi,\phi_2)$ is the distance between two points in moduli space.

Hence the $2^{nd}$ law of species thermodynamics says that the number of species must not decrease
\begin{equation}
\delta N_{sp}\geq 0.
\end{equation}
According to the arguments elaborated upon in this work, this means that the motion in moduli space of KK towers goes in the direction of decompactification, i.e.~towards large volume: $\delta {\cal V}(\phi)\geq 0$. Instead, for string towers the motion goes towards weak string coupling, i.e.~$\delta g_s(\phi)\leq 0$.
For the variation of cosmological constant we get by the use of the ADC that $\delta \Lambda_{cc}(\phi)\leq 0$.

Let us comment more on this second law.  
Since the species  energy $E_{sp}$ scales as $E_{sp}\simeq {\cal S}_{sp}^{\frac{d-3}{d-2}}$, we get that $\delta E_{sp}\geq0$.
As shown before (see also the appendix \ref{AppendixA}), for KK tower of degeneracy $n$, the energy differences scale as $\Delta E\sim {\cal S}_{sp}^{\frac{2-(n+d)}{n(d-2)}}$, so the energies differences must not increase, namely $\delta(\Delta E)\leq 0$.
Finally, for concreteness, let us focus on only one compact dimension. Then, the ground state tower mass scales as $m_0\simeq ({\cal S}_{sp})^{\frac{1-d}{d-2}}$, and we obtain that $\delta m_0\leq 0$.
This is an insight on what the Swampland Distance Conjecture predicts. The $2^{nd}$ law of species thermodynamics is related to the Swampland Distance Conjecture in the sense that, for the strict inequality in \eqref{2ndlaw}, when moving towards the boundary of the moduli space one gets an increasing number of species with lower and lower masses. In this sense, the Swampland Distance Conjecture implies the $2^{nd}$ law of species thermodynamics. On the other hand, when considered as a strict inequality, this $2^{nd}$ law combined with the BHEDC, leads to the Swampland Distance Conjecture. 
The situation in which the inequality is saturated, instead, is more subtle and it probably requires more information about the dynamics on the moduli space. In our present understanding, the $2^{nd}$ law does not forbid adiabatic motion in which the species scale remains constant. 
Besides, let us notice that we can also express the field derivative of $m_0(\phi)$ as (derivatives are expressed with a prime)
\begin{equation}
\frac{m_0^{'}}{m_0}\simeq -\biggl({\frac{d-1}{d-2}}\biggr) \frac{\mathcal{S}_{sp}^{'}}{\mathcal{S}_{sp}}\, .
\end{equation}

It is important to note again that, differently from the black hole thermodynamic laws, there is no time parameter a priori. Here the role of the time is substituted by real scalar quantities, namely the (scalar fields) moduli $\phi$.

Furthermore, if two towers of species coalesce, the final species scale is always less then the minimum of the initial species scales
\begin{equation}\label{coalesceSpecies}
    \Lambda_{sp_1 + sp_2} \leq \min(\Lambda_{sp_1}, \Lambda_{ sp_2})\, .
\end{equation}

This second law of species thermodynamics is slightly stronger than the corresponding thermodynamic law.
This means that the coalesce of two species towers produce another species towers such that the latter has a quantum gravity cut-off which is at most the minimum of the two species scales.

\item {\bf Third law of species thermodynamics.}

It is impossible by any physical process to reduce the species temperature $T_{sp}$ to zero by a finite sequence of operations. 

This law establishes that we cannot reach the boundary of the moduli space, i.e.
\begin{align*}
    \forall N \in \mathbb{N},\; \exists \epsilon > 0,  \; &\textrm{ s.t for every sequence of moduli }  \{\phi_N\} \in \mathcal{M}
    \end{align*}
    one has
    \begin{align*}
    \min_{\phi \in \partial \mathcal{M}}\Delta(\phi_N, \phi) > \epsilon.
\end{align*}

In other words it means that the boundary of the moduli space is at infinite distance. In particular, the weak coupling limiting values ${\cal V} \rightarrow \infty$ for KK towers and $g_s \rightarrow 0$ for the string tower cannot be reached.
It means that the complete break down of effective field theory with vanishing UV cutoff scale $\Lambda_{sp} \rightarrow 0$ is at infinite distance and can never be reached. In the same way, the limit of vanishing cosmological constant $\Lambda_{cc} \rightarrow 0$
cannot be reached by a finite sequence of operations.

\end{itemize}

\section{Summary and Outlook}

In this paper, we proposed that the entropy ${\cal S}_{sp}$ of particle species coupled to quantum gravity is identical to the Bekenstein-Hawking entropy
of a black hole with radius $R_{sp} = \Lambda_{sp}^{-1}$, i.e. ${\cal S}_{sp}\simeq (R_{sp})^{d-2}$. For a tower of species with increasing masses up to  $\Lambda_{sp}$, the species entropy is approximately given by the number of species as ${\cal S}_{sp}\simeq N_{sp}$.  
This definition of species entropies agrees with the entropy of the smallest possible black holes in the effective theory, when it is calculated by first reducing the size of the black hole and the by increasing the entropy as much as possible.
Thus, the leading order expressions for the entropy in terms of $N_{sp}$ obtained in this way are valid near the boundary of the moduli space.
Additional information about the species entropy inside the moduli space can be obtained from the topological string partition function $F_1$ or from a duality invariant notion of species scale.

Instead, for $N_0$ light or massless species, the species entropy receives multiplicative log-corrections and is of the order ${\cal S}_{sp}\simeq N_{0}\log N_0$. Concretely, the entropy formula can be derived by a perturbative calculation, again in the limit of large $N_0$.

The species temperature can be deduced from the temperature of non-extremal, minimal black holes, giving $T_{sp} = {\cal S}_{sp}^{-1/(d-2)}$. We also introduced the species energy to be $E_{sp}={\cal S}_{sp}^{(d-3)/(d-2)}$. 
All of these definition agree with what is expected from standard black hole thermodynamics.
With this information at hand, we could set up the species thermodynamics, which have interesting implications for the motion of the species variables in moduli space. In particular the $2^{nd}$ law of species thermodynamics and the species distance conjecture are closely related to some of the basic swampland conjectures, such as the swampland Distance Conjecture: they imply an increasing tower of massless species when moving towards the boundary of the moduli space. According to the $3^{rd}$ law of species thermodynamics, the  boundary can however never be reached, for it lies at infinite distance.
To summarize, the framework of species thermodynamics provides a new thermodynamic and entropic point of view on these swampland conjectures and ideas. 

In case one can identify one direction in moduli space with the real time coordinate, as for
backgrounds with time dependent dilaton or moduli fields, the species thermodynamics may have also important implications for cosmology or the time dependence of parameters in the effective action. As time increases, one gets driven towards weaker and weaker couplings, while the cosmological constant gets smaller and smaller and reaches zero at an infinite amount of time. More precisely, we expect this behavior to hold true in asymptotic regions of the moduli space, where our thermodynamic picture is reliable. Nevertheless, one can contemplate the possibility that the species scale grows with time for a (short) period in the bulk of the moduli space, before assuming its decreasing behavior asymptotically. In general, what happens in the interior of the moduli space is an interesting open question, to which we hope to come back in the future.

One open question is the role of Hawking radiation in the context of species thermodynamics.  Like for black holes, one could expect that Hawking radiation leads to a decrease of the species entropy, i.e.~to a decay of the tower of species 
with a decay spectrum determined by the species temperature $T_{sp}$,  in semi-classical approximation. 
It would be very interesting to explore the physical mechanism of the species Hawking radiation, 
and also to see how species Hawking radiation might possibly related to the decay of dark KK gravitons in the dark dimension scenario
\cite{Gonzalo:2022jac,Anchordoqui:2022tgp}.

Another interesting open problem could be a possible connection between the species entropy and holographic entanglement entropy of space-time.

Finally, on the more mathematical side, it would be interesting to connect the laws of species thermodynamics with flow equations, like Ricci flow, which were
considered in the context of the swampland in  \cite{Kehagias:2019akr}  and in additional series of papers 
\cite{DeBiasio:2020xkv,Velazquez:2022eco,DeBiasio:2022nsd,DeBiasio:2022zuh}.

\section*{Acknowledgments}

We thank Ivano Basile, Ralph Blumenhagen, Severin L\"ust, Miguel Montero and Cumrun Vafa for useful discussions. The work of N.C.~is supported by the Alexander-von-Humboldt foundation.
The work of D.L.~is supported by the Origins
Excellence Cluster and by the German-Israel-Project (DIP) on Holography and the Swampland.

\appendix
\section{General calculation of the species energy}\label{AppendixA}
In order to provide a general proof of \eqref{EnergyRel}, let us consider a generic KK tower. As is known, the mass spectrum grows polynomially and it is defined as \cite{Castellano:2021mmx}
\begin{equation}
    M_k \simeq k^{1/n} m_{KK},
\end{equation}
where $n$ is the degeneracy of states, and $m_{KK} = \Delta E$ is the minimal possible mass in the tower.
Let us denote with $N$ the last species in the tower. This means that the KK mass calculated at $k=N$ gives the species scale
\begin{equation}\label{PolynDeg}
    \Lambda_{sp} = M_N \simeq N^{1/n} m_{KK} \implies m_{KK} = \frac{\Lambda_{sp}}{N^{1/n}}.
\end{equation}
In order to calculate the species energy, we need to sum over every mass present in the tower
\begin{equation}
    E_{sp} \simeq \sum_{k=0}^N k^{1/n} m_{KK} \sim m_{KK} \, H_N^{(-1/n)} \equiv  m_{KK} \left[-\zeta\left( -\frac{1}{n}, N+1\right) + \zeta\left(-\frac{1}{n}\right) \right]
\end{equation}
where $\zeta(s,q)$ is the Hurwitz Zeta function, and $H_n^{a}$ is the generalized harmonic function.

We are interested in the large $N$ limit, in which our whole analysis takes place. In such limit, the behavior of the Hurwitz Zeta function allows us to write the following asymptotic relation
\begin{equation}
    E_{sp} = m_{KK} N^{1/n} \left(\frac{n}{n+1} N +\frac{1}{2} + \mathcal{O}\left(\frac{1}{N} \right) \right).
\end{equation}
Substituting $m_{KK}$ and restoring the Planck mass, we obtain
\begin{equation}\label{GeneralEnergyRel}
    E_{sp} \simeq \Lambda_{sp} \frac{n}{n+1} N = \frac{n}{n+1} \Lambda_{sp}^{3-d} M_P^{d-2},
\end{equation}
which is indeed the general proof of the formula \eqref{EnergyRel}.
In particular the string tower limit is obtain for $n \rightarrow \infty$. As we can see from the result \eqref{GeneralEnergyRel}, the relation between the species energy and the species scale does not change.

Let us prove, in a more formal way, this result for the string tower.
For a string tower the states degeneracy $d_n$, at large excitation level number $n$, is not polynomial\eqref{PolynDeg}, but (approximately) exponential: $d_n \sim e^{\sqrt{n}}$. This means that, when counting the number species under the UV cutoff, one must take this effect into account. Let us consider the large $N_{sp}$ limit and proceed as above by setting
\begin{eqnarray}
    \Lambda_{sp}^{d-2} = \frac{M_P^{d-2}}{N_{sp}},
\end{eqnarray}
with
\begin{equation}
    N_{sp} = \sum_{k=0}^{N_{s}} d_k \sim \sum_{k=0}^{N_{s}}e^{\sqrt{k}} \simeq \sqrt{N_{s}} e^{\sqrt{N_{s}}},
\end{equation}
 and $N_s$ is the maximum excitation level such that the mass is below the quantum gravity cut-off $\Lambda_{sp}$.
 Indeed, we can define the species scale also through the string mass spectrum
\begin{equation}
    \Lambda_{sp} = m_{N_{s}} \simeq \sqrt{N_{s}} M_s .  
\end{equation}
From this definition, one can calculate $N_{s}$ in terms of the string mass $M_s$ via the $(-1)$-branch of the Lambert function $W_{-1}$\cite{Castellano:2022bvr},
\begin{equation}
    \sqrt{N_{s}} \simeq \log{\frac{M_P}{M_s}}.
\end{equation}
This allows us to write the species scale in term of $M_s$ as
\begin{equation}
    \Lambda_{sp} \simeq M_s \log{\frac{M_P}{M_s}}.
\end{equation}
Now, we can calculate the species energy as the weighted sum of the string masses
\begin{equation}
    E_{sp} \simeq M_s \sum_{k=0}^{N_{s}} d_k \sqrt{k} \sim  M_s  \sum_{l=0}^\infty \frac{1}{l!} \sum_{k=0}^{N_{s}} k^{\frac{l+1}{2}} 
\end{equation}
Finally, as before, we use the asymptotic behavior of the Hurwitz Zeta function in the large $N_{s}$ limit 
\begin{equation}
    E_{sp} \simeq M_s \sum_{l=0}^\infty \frac{1}{(l+1)!} N_{s}^{\frac{l+1}{2}} N_{s} \sim M_s N_{s} e^{\sqrt{N_{s}}}  = M_s \sqrt{N_{s}} N_{sp},
\end{equation}
hence
\begin{equation}
    E_{sp} \simeq \Lambda_{sp} \Lambda_{sp}^{2-d} M_P^{d-2} = \Lambda_{sp}^{3-d} M_P^{d-2}.
\end{equation}

This calculation shows that the thermodynamic quantities of the species tower are fairly independent from the microscopic nature of the tower itself, such as degeneracy of the states, compactification details and so on. In fact, as we argued for in this paper, the 
Schwarzschild black holes associated to the species scale can be represented by a generic tower. In other words, this result explain that the thermodynamic behavior of the species is rather universal and independent from many details of the microscopic point of view.

\end{document}